\documentclass[nofootinbib,notitlepage,superscriptaddress,10pt,aps,pra,twocolumn]{revtex4-1}

\usepackage[utf8x]{inputenc}
\usepackage{amsmath}

\usepackage{amssymb}
\usepackage{bbm}
\usepackage{graphicx}

\usepackage[normalem]{ulem}

\begin{document}

\title{Misleading  inferences from discretization of empty spacetime:\\ Snyder-noncommutativity case study}

\author{Giovanni AMELINO-CAMELIA}
%\email{Giovanni.Amelino-Camelia@roma1.infn.it}
\affiliation{Dipartimento di Fisica, Universit\`a di Roma ``La Sapienza", P.le A. Moro 2, 00185 Roma, Italy}
\affiliation{INFN, Sez.~Roma1, P.le A. Moro 2, 00185 Roma, Italy}

\author{Valerio ASTUTI}
\affiliation{Dipartimento di Fisica, Universit\`a di Roma ``La Sapienza", P.le A. Moro 2, 00185 Roma, Italy}
\affiliation{INFN, Sez.~Roma1, P.le A. Moro 2, 00185 Roma, Italy}

\begin{abstract}
Alternative approaches to the study of the quantum-gravity problem are handling the role of spacetime
very differently. Some are focusing on the analysis of one or another
novel formulation of ``empty spacetime", postponing to later stages the
introduction of particles and fields, while other approaches assume that spacetime should only be an emergent entity.
We here argue that recent progress
in the covariant formulation of  quantum mechanics suggests that empty spacetime
is not physically meaningful.
We illustrate our general thesis in the specific context of the noncommutative Snyder  spacetime,
which is also of some intrinsic interest, since hundreds of studies were devoted to its
analysis.
We show that empty Snyder spacetime, described in terms of a suitable kinematical Hilbert space, is discrete,
but this is only a formal artifact: the discreteness leaves no trace on
the observable properties of particles on the
physical Hilbert space.\\
\end{abstract}

\maketitle

\section{Introduction}
Nearly all approaches to (and analyses of)\ the quantum-gravity problem lead to the expectation that spacetime itself should
acquire quantum properties, when probed at the Planck
scale ($\simeq 10^{28}eV$). Due to a dearth
of experimental clues, a variety of approaches are being
developed toward new pictures of spacetime. Two broad categories
can
be identified on opposite sides. Some approaches start off without a spacetime (see, {\it e.g.}, Refs.\cite{AmelinoCamelia:2011bm,Konopka:2008hp}) and expect spacetime to emerge from the picture at some effective-theory level, with non-classical properties induced by the emergence mechanism. Other approaches introduce specifications of non-classical properties of spacetime from the beginning at the level of the structural postulates of the theory
(see, {\it e.g.}, Refs.\cite{Doplicher:1994tu,Majid:1994cy}). We are here concerned
with this latter class of studies and specifically the subclass of cases in which the development
of the theory starts off with a characterization of ``empty spacetime", {\it i.e.} studies focusing merely on the analysis of the nonclassical geometry
of spacetime. In such approach one postpones the introduction of particles to later stages of the development of the research programme, also hoping  that the analysis of the geometry of empty spacetime could provide a reliable intuition on the general properties of theories in such a spacetime.

Our main concerns originate from the fact that
recent advances~\cite{Halliwell:2000yc,Gambini:2000ht,Reisenberger:2001pk} in the manifestly-covariant formulation of quantum mechanics suggest that the structure of spacetime does affect directly  the properties of the unphysical ```kinematical Hilbert space", but has a more indirect role in the properties which are really of interest, coded on the ``physical Hilbert space".

We show that these concerns should not go overlooked by analyzing
one of the most studied candidates for quantum spacetime:
the Snyder spacetime \cite{Snyder} characterized by noncommutativity of coordinates of the form
\begin{eqnarray}
[x^{\mu},x^{\nu}] = i\lambda^2 M^{\mu \nu}
\label{snycomm}
\end{eqnarray}
where $\lambda$ is the length scale of Snyder noncommutativity
and $M^{\mu \nu}$ denotes the Lorentz generators.

This noncommutative Snyder spacetime has been studied extensively,
especially in relation to the possibility of using it to model
physical pictures in which (some of)\ the spacetime coordinates
are discretized. The analysis we here report of Snyder noncommutativity
within the manifestly-covariant formulation of quantum mechanics
turns out to provide a prototypical example of how our concerns
can get very serious. We find that at the level of analysis involving the unphysical kinematical Hilbert space Snyder's proposal does indeed involve a discretization of coordinates,
but this is only a formal artifact: the discreteness leaves no trace on
the observable properties of particles on the physical Hilbert space.

\section{Preliminaries on the covariant formulation of quantum mechanics}
In spite of the sizable literature devoted to the Snyder spacetime
some of the aspects we here uncover
had been completely overlooked.
Presumably the main reason for this is that the structure of the  Snyder spacetime involves a law of noncommutativity
between time coordinate and spatial coordinates.
In the traditional formulation of quantum mechanics such a noncommutativity
cannot be faithfully accommodated because time and space are handled in
profoundly different manners: the spatial coordinates are described by operators
 whereas the time coordinate is just
an external evolution parameter. This apparent obstacle for actually doing
quantum mechanics in the Snyder spacetime ended up directing most of the associated research
effort toward studies of mathematical properties of the Snyder coordinates, which should be viewed as properties
of the ``empty" Snyder spacetime, in a sense that will become clearer as we go along.

We here take as starting point the fact that the standard formulation of quantum mechanics should be viewed merely
as a sort of gauge-fixed version of the more general and more powerful manifestly-covariant
formulation of quantum mechanics.
This formulation of quantum mechanics has  progressed
significantly over the last decade (see, {\it e.g.}
Refs.~\cite{Halliwell:2000yc,Gambini:2000ht,Reisenberger:2001pk}), rendering
increasingly clear how spatial and time coordinates there play
exactly the same type of role.
In the standard case (commutative spacetime) spatial and time coordinates are well-defined
operators~\cite{Reisenberger:2001pk} on
a ``kinematical Hilbert space", which is
the space of square-integrable functions $L^2(\mathbb{R}^4,dq^0dq^1dq^2dq^3)\sim L^2(\mathbb{R}^2,dp^0dp^1dp^2dp^3)$,
with canonical commutation relations
\begin{equation}
[q^{\mu}, p_{\nu}] = i \delta^{\mu}_{\nu} \,\, , \,\,\,\,\,\,[q^\mu, q^\nu]=0 \,\, , \,\,\,\,\,\,[p_\mu, p_\nu]=0 \, .
\end{equation}
Here we denote with $q^{\mu}$ the spacetime coordinates, while the $p_\mu$ are of course the conjugate momenta.

Time is not an evolution parameter because in the covariant formulation of quantum mechanics
there is no ``evolution":
dynamics is codified in a constraint, just in the same sense
familiar for the covariant formulation of classical mechanics (see, {\it e.g.}, Ref.\cite{henneaux}).
For example, for free particles the constraint simply enforces on-shellness,
and the ``reduced" Hilbert space obtained from the kinematical Hilbert space
by enforcing on-shellness is the so-called ``physical Hilbert space".

The physical content of the theory resides in the properties of self-adjoint operators
on the physical Hilbert space. Through these one recovers the information on physical evolution
of systems from the formal setup which involves a pure-constraint Hamiltonian.
This will be crucial for our analysis of the physical implications
of Snyder noncommutativity, so we devote a few equations
to showing some aspects  of its implementation (more details
in Refs.~\cite{Halliwell:2000yc,Gambini:2000ht,Reisenberger:2001pk}, and Refs.~\cite{fuzzy1,fuzzy2} for a recent application to noncommutative spacetimes).
For free particles one has that states of the physical Hilbert space must comply
with the requirement
$$ {\cal H}\psi =[p_0^2 - p_1^2 - m^2]\psi =0$$

A convenient strategy for implementing the constraint
is based~\cite{Reisenberger:2001pk}
on introducing a corresponding scalar product on
the physical Hilbert space, such that
different kinematical states
are projected on the same physical state. Specifically
one adopts the following scalar product:
\begin{equation}
\langle \phi |\psi\rangle = \int d\mu(p) \,\, \delta({\cal H}) \,
\Theta(p_0) \,\, \overline{\phi} ( p ) \, \psi ( p )
\end{equation}
where $\Theta(p_0)$  specifies a restriction~\cite{Reisenberger:2001pk}
 to positive-energy solutions of the on-shellness constraint.

It is important to stress
that the spacetime coordinates $q^\mu$ are self-adjoint operators
on the kinematical Hilbert space, but they are not self-adjoint operators on the physical
Hilbert space, since $[q^\mu,{\cal H}]\neq 0$. So the physical properties of the theory
cannot be captured by properties of the operators $q^\mu$. They are coded  in properties
of combinations of the operators $q^\mu$ and $p_\mu$ which commute
with the Hamiltonian constraint. An example of such physical observables
are
the Newton-Wigner operators
\begin{equation}
A_i = q_i - \frac{p_i}{p_0}q_0 + i\frac{p_i}{2p_0^2}
\end{equation}
which indeed are self-adjoint on the physical Hilbert space (in particular $[A_i ,{\cal H}] = 0$).

How is this regaining us a picture of evolution from the ``frozen" pure-constraint setup?
This is easily seen by observing that, for example,
the expectations of the operators $p_i$ and $A_i$
in a state $\psi$ of the physical Hilbert space,
$$<\psi | p_i |\psi >={\bar p}_i \,\, , \,\,\,\,\,\,
 <\psi | A_i |\psi >=a_i \, ,$$
are meaningful physical properties. These are just expectations of observables \uline{in a given (frozen) state} that satisfies the Hamiltonian constraint, but contain information on the \uline{evolution} of the system. One can easily see this by contemplating the interpretation of the Newton-Wigner operator in classical mechanics:
the state $\psi$ is placing the particle on a worldline (a fuzzy worldline in quantum mechanics) which in the classical limit is such that
$$q_i = a_i + \frac{{\bar p}_i}{p_0({\bar p}_i)}q_0 $$
(because of the Hamiltonian constraint, on the physical Hilbert space $p_0$ must be viewed as constrained in terms of the $p_i$).
So the particle is moving with standard velocity $ {\bar p}_i/{p_0({\bar p}_i)}$
and intercepts the $q^i$ axis at $q^i = a^i $.

\section{Discreteness of empty Snyder spacetime}

We reminded our readers in the previous section of the fact that even the standard (commutative) spacetime coordinates are not self-adjoint operators on the physical Hilbert space. They are not observables. They are however well-defined operators on the kinematical Hilbert space, well suited
for characterizing the geometric structure of ``empty" Snyder spacetime (there
are evidently no physical particle on the kinematical Hilbert space).
Then any proposal of noncommutativity of spacetime coordinates must be formulated first on the kinematical Hilbert space, keeping in mind however that the observable manifestations of coordinate noncommutativity can only be its implications for the properties of self-adjoint operators on the physical Hilbert space. In this section we shall see how the Snyder noncommutativity admits a satisfactory description on a kinematical Hilbert space,
which in particular involves a discretization of spatial coordinates.

\subsection{Kinematical-Hilbert-space representation }
In giving a representation for the noncommutative Snyder coordinates on a kinematical Hilbert space, we keep the notation $p_\mu$, $q^\mu$. So,
in particular, we still have
that $[q^{\mu}, p_{\nu}] = i \delta^{\mu}_{\nu}$. The difference is that
now the $q^{\mu}$ will not be interpreted as spacetime coordinates,
but only as auxiliary operators useful for the description of
Snyder's noncommutative coordinates $x^\mu$. One easily sees that
  a satisfactory representation of Snyder noncommutativity is obtained by posing:
\begin{equation}
x^{\mu} = q^{\mu} - \lambda^2 p^{\mu}(p\cdot q) \, .
\label{represnyder}
\end{equation}
Indeed we have that
\begin{eqnarray}
[x^{\mu},x^{\nu}] &=& [q^{\mu} - \lambda^2 p^{\mu}(p\cdot q), q^{\nu} - \lambda^2 p^{\nu}(p\cdot q)] =  \nonumber \\
&=& -\lambda^2 \left([p^{\mu}(p\cdot q),q^{\nu}]+[q^{\mu},p^{\nu}(p\cdot q)]\right) + \nonumber \\
&+& \lambda^4 [p^{\mu}(p\cdot q),p^{\nu}(p\cdot q)]= \nonumber \\
&=& i\lambda^2 \left(p^{\mu}q^{\nu} - p^{\nu}q^{\mu}\right) = \nonumber \\
&=& i\lambda^2 \left(p^{\mu}x^{\nu} - p^{\nu}x^{\mu}\right) = i\lambda^2 M^{\mu \nu}
\end{eqnarray}
It is also easy to verify that in order for the $x^{\mu}$ to be
hermitian operators the scalar product on the kinematical Hilbert space
must involve the following  $\lambda$-deformed
integration measure ($p^2$ denotes as usual $p_\mu p^\mu$)
\begin{equation}
d\mu(p) = \frac{d^4p}{(1-\lambda^2 p^2)^{5/2}}
%\approx d^4p \left(1+\frac{5}{2}\lambda^2 p^2\right)
\, .
\label{measure}
\end{equation}
The coordinates  $x^{\mu}$ are self-adjoint operators in this Hilbert space. We shall soon study their  spectra, but first let us be explicit about the evaluation of expectation values. In the momentum-space representation, with states $\psi(p)$ on the kinematical Hilbert space, one has for example that the expectation of  $x^{\mu}$ is given by
\begin{equation}
\langle x^{\mu} \rangle = i \frac{\int d\mu(p)\, \overline{\psi(p)}
\left(g^{\mu\rho} - \lambda^2 p^{\mu}p^{\rho}\right)
\frac{\partial}{\partial p^{\rho}}\, \psi(p)}{\int d\mu(p) \left|\psi(p)\right|^2}
\end{equation}
where $d\mu(p)$\ is the integration measure of (\ref{measure})
and $\left(g^{\mu\rho} - \lambda^2 p^{\mu}p^{\rho}\right)
i\frac{\partial}{\partial p^{\rho}}$ is the momentum-space representation
of $x^{\mu}$, on the basis of (\ref{represnyder}).

It is also easy to verify that this construction (as stressed already in Snyder's original work \cite{Snyder}) preserves Lorentz symmetry. This is essentially due to the fact that Snyder's commutation
relations (\ref{snycomm}) are Lorentz invariant and the measure of
integration (\ref{measure}) is Lorentz invariant.

\subsection{Discreteness}
As anticipated in Snyder's original paper, and more recently established in detail in Ref.\cite{stern}, Snyder's spacetime leads to a picture of  space in cartesian coordinates as a (quantum) cubical lattice of lattice spacing $\lambda$ (for $\lambda$ real number with dimensions of length, as we are also here assuming),
 while time remains a continuous variable.
These results have been established by seeking formal Hilbert-space representations of the Snyder algebra, but the question remained so far concerning the physical interpretation of these results. As already stressed above within the standard formulations of quantum theory there is no room for such a Hilbert space, on which in particular the time coordinate is described by an operator. By exploiting recent progress in the manifestly covariant formulation of
quantum mechanics we gave in the previous subsection the correct interpretation
in physics for the relevant Hilbert space, as the kinematical Hilbert space. We can therefore quickly verify that on our kinematical Hilbert space Snyder's spatial coordinates indeed have a discrete spectrum.

For this purpose it is sufficient to notice that in the construction given in the previous subsection the spatial coordinates and the angular momentum generators form an $SO(4)$ algebra. This is easily seen by posing
\begin{equation}
L_{ij} = M_{ij}  \quad, \quad L_{i4}= \frac{x_i}{\lambda}  \quad, \quad L_{AB} = - L_{BA} \, ,
 \nonumber
\end{equation}
which indeed, also in light of (\ref{snycomm}), reproduce the commutation relations for the $SO(4)$ algebra:
\begin{equation}
[L_{AB}, L_{CD}] = i (\delta_{AC}L_{BD} - \delta_{BC}L_{AD} - \delta_{AD}L_{BC} + \delta_{BD}L_{AC})
\nonumber
\end{equation}
This observation shows that one can find a basis for our kinematical Hilbert space which is built of representations of this algebra. The implications of this are best seen by introducing the six operators
\begin{equation}
A_i = \frac{1}{2} \left(L_i + \frac{x_i}{\lambda}\right) \,\, , \,\,\,\,\,\,
 B_i = \frac{1}{2} \left(L_i - \frac{x_i}{\lambda}\right) \, ,
\end{equation}
with $L_i = \epsilon_{ijk}M^{jk}$; this allows us to decouple the relevant $SO(4)$ algebra in two $SU(2)$
algebras: $[A_i,A_j]=i\epsilon_{ijk}A_k$, $[B_i,B_j]=i\epsilon_{ijk}B_k$ and $[A_i,B_j]=0$, with $A_iA_i = B_iB_i$.

Then we can have basis states labeled by eigenvalues of $A_3$, $B_3$, and the casimir $A_iA_i$:
\begin{equation}
A_3|j,m_A,m_B\rangle = m_A |j,m_A,m_B\rangle
\end{equation}
\begin{equation}
B_3|j,m_A,m_B\rangle = m_B |j,m_A,m_B\rangle
\end{equation}
\begin{equation}
A_iA_i|j,m_A,m_B\rangle = j(j+1) |j,m_A,m_B\rangle
\end{equation}
On these basis states we obtain for the $x_3$ coordinate, which can be written
 as $x_3 = \lambda \left(A_3 - B_3 \right)$, that
 \begin{equation}
x_3|j,m_A,m_B\rangle = \lambda (m_A - m_B)|j,m_A,m_B\rangle
\end{equation}
The eigenvalues $m_A$, $m_B$ are integer numbers, so the eigenvalues of the $x_3$ coordinate form a one dimensional lattice of spacing $\lambda$.
The same reasoning applies of course also to the other spatial
 coordinates, $x_1$ and $x_2$,
 so we can conclude that the spatial part of the Snyder spacetime
 is a 3-dimensional quantum lattice (but states in the kinematical Hilbert space which are eigenstates of one coordinate are not eigenstates of the other two coordinates),
  with infinitely
 degenerate eigenvalues.

\section{Futility of discretization of Snyder spacetime}
On the kinematical Hilbert space of quantum mechanics, Snyder's spacetime is mainly characterized by the introduction of a deformed scalar
product, involving the deformed integration
measure (\ref{measure}), and by the discretization of spatial coordinates.
But there are no physical observables on the kinematical
Hilbert space \cite{Halliwell:2000yc,Gambini:2000ht,Reisenberger:2001pk}.
Physical observables can only be introduced on the physical Hilbert space,
after enforcing a suitable Hamiltonian constraint. The discretization
of Snyder's spacetime could be meaningful only to the extent that it
would leave a trace in such observables on the physical Hilbert space. Let us then consider the propagation of particles in Snyder's spacetime.
As shown above Snyder's proposal leaves Lorentz symmetry unaffected, so
in order to describe propagation of particles we need
to enforce the standard Hamiltonian constraint
$$ [p_0^2 - p_1^2 - m^2]\psi =0$$
The kinematical Hilbert space is ``reduced" to the physical
Hilbert space of states $\psi$ that satisfy this constraint.\ And
we start by noticing that the modification of the scalar product affecting
the kinematical Hilbert space trivializes when restricted to states
in the physical Hilbert space
\begin{equation}
d\mu(p) = \frac{d^4p}{(1-\lambda^2 p^2)^{5/2}} \longrightarrow
d\mu(p) = \frac{d^4p}{(1-\lambda^2 m^2)^{5/2}}\, ,
\nonumber
\end{equation}
This apparent deformation of the measure of integration just amounts to  multiplication by a constant (the mass $m$ is fixed on the entire physical Hilbert space once and for all). It therefore gets reabsorbed in the normalization of the states, and is completely irrelevant.

This observation about scalar products takes us one step closer to establishing
the futility of the results obtained for Snyder's spacetime at the level
of the kinematical Hilbert space.
The possibility of discretization of the coordinates $x_j$ is not even
a meaningful possibility on the physical Hilbert space, since
evidently the $x_j$
do not commute with the Hamiltonian constraint operator
and therefore cannot be \cite{Halliwell:2000yc,Gambini:2000ht,Reisenberger:2001pk}
observables on the physical Hilbert space.
What one could hope for is for the discretization of the $x_j$ on
the kinematical Hilbert space to manifest itself under a different
disguise  as some property of observables on the physical Hilbert space.
But we find that this is not the case. To see this let us start
from
the observables on the physical Hilbert space for free propagating particles which have been most considered in the relevant literature. For the undeformed theory ($\lambda = 0$)  the operators one usually considers are \cite{freid}
 the operators
${\cal X}^{\mu} = q^{\mu} - p^{\mu}(q\cdot v)/(p\cdot v) + h.c.$
where $v$ is a real-valued, time-like vector that parametrizes these operators that commute with the Hamiltonian constraint. This family of operators of course also includes, for $v^\mu = \delta^{\mu 0}$, the Newton-Wigner operator which we already discussed above.

We notice that for the Snyder spacetime one does get good self-adjoint
operators on the physical Hilbert space (commuting with
the Hamiltonian-constraint operator)
by taking these standard ${\cal X}^{\mu}$ and replacing the $q^{\mu}$
with the Snyder coordinates $x^{\mu}$:
\begin{equation}
{\cal X}_{{[\lambda]}}^{\mu} = x^{\mu} - \frac{p^{\mu}}{p\cdot v}x\cdot v + h.c.
\end{equation}
It is therefore meaningful to ask whether  the discretization of the $x_j$ on the (unobservable)
kinematical Hilbert space
leaves any trace on properties of these ${\cal X}_{{[\lambda]}}^{\mu}$
observables on the physical Hilbert space. The answer
is no, as one can easily see by recalling that the
Snyder coordinates admit the representation
$x^{\mu} = q^{\mu} - \lambda^2p^{\mu} (p\cdot q)$, from which
it follows that
\begin{eqnarray}
{\cal X}^{\mu}_{[\lambda]} &=& x^{\mu} - \frac{p^{\mu}}{p\cdot v}x\cdot v + h.c.\nonumber \\
&=& q^{\mu} - \lambda^2p^{\mu} (p\cdot q) - \frac{p^{\mu}}{p\cdot v} (q^{\nu} - \lambda^2p^{\nu} (p\cdot q))v_{\nu} + h.c.\nonumber \\
&=& q^{\mu} - \frac{p^{\mu}}{p\cdot v} (q\cdot v) + h.c. = {\cal X}^{\mu}
\label{main}
\end{eqnarray}
The definition of the $\cal X^{\mu}_{[\lambda]}$ does involve the Snyder
deformation scale $\lambda$ but it turns out that
actually they   are $\lambda$-independent.
The $\cal X^{\mu}_{[\lambda]}$ on the physical Hilbert space
are insensitive to the discretization of the $x^{\mu}$ on the kinematical Hilbert space. They are completely independent of the deformation scale $\lambda$.

On the basis of results obtained in the previous literature on the manifestly covariant formulation of quantum mechanics the observables  $\cal X^{\mu}_{[\lambda]}$ are particularly significant since they come as close as one can get  \cite{freid}
to the notion of a spacetime coordinate on the physical Hilbert space.\ A mentioned special cases of the  $\cal X^{\mu}_{[\lambda]}$  are the Newton-Wigner operators, to which then evidently our result (\ref{main}) applies:

\begin{eqnarray}
A^i_{[\lambda]} &=& x^{i} - \frac{p^i}{p^0}x^0 + h.c. =\nonumber \\
&=& q^{i} - \lambda^2p^{i} (p\cdot q) - \frac{p^{i}}{p^{0}} (q^{0} - \lambda^2 p^{0} (p\cdot q)) + h.c. = \nonumber \\
&=& q^i - \frac{p^i}{p^0} + i\frac{p^i}{2(p^0)^2} = A^i
\nonumber
\end{eqnarray}

While these results for the observables  $\cal X^{\mu}_{[\lambda]}$,
and particularly the $A^i_{[\lambda]}$, already give a rather strong
intuition for the mechanism which is at play, it is noteworthy that
we are in position to show, with even greater generality, that observables on the physical Hilbert space are unaffected by the $\lambda$-deformation introduced by Snyder's spacetime.
We do not expect to loose much generality by focusing on quantum observables
which admit a smooth classical limit, and by restricting our proof to the
case of a  1+1 dimensional Snyder spacetime (additional space dimension just replicate what we expose for the one spatial dimension of the
1+1 dimensional Snyder spacetime). We start by noticing that in order to commute with the Hamiltonian constraint such an observable $f(p,q)$
should be such that
\begin{equation}
p^{\mu}\frac{\partial f}{\partial q^{\mu}} =0
\end{equation}
So if  $f(p,q)$ is an observable on the physical Hilbert space
it must be the case that the
vector $\frac{\partial f}{\partial q^{\mu}}$ is orthogonal to
the momentum $p^{\mu}$.
We can conclude that the vector function $\frac{\partial f}{\partial q^{\mu}}$ must be of the form:
\begin{equation}
\frac{\partial f}{\partial q^{\mu}} = f'(p,q)\epsilon_{\mu \nu}p^{\nu}
\end{equation}
Taking another derivative and imposing the symmetry between partial derivatives we have:
\begin{equation}
\frac{\partial^2 f}{\partial q^{\mu}\partial q^{\rho}}(q,p) = \frac{\partial f'(q,p)}{\partial q^{\rho}}\epsilon_{\mu \nu}p^{\nu} = \frac{\partial f'(q,p)}{\partial q^{\mu}}\epsilon_{\rho \nu}p^{\nu}
\end{equation}
From this it is easy to establish that
\begin{equation}
\frac{\partial f'(q,p)}{\partial q^{\rho}}p^{\rho} = 0
\end{equation}
and therefore, just
like $\frac{\partial f}{\partial q^{\mu}}
= f'(p,q)\epsilon_{\mu \nu}p^{\nu}$, one has that
\begin{equation}
\frac{\partial f'(q,p)}{\partial q^{\rho}} = f''(q,p)\epsilon_{\rho \nu}p^{\nu}
\end{equation}
The argument can be easily iterated, to find that\begin{equation}
\frac{\partial^n f(q,p)}{\partial q^{\mu}\partial q^{\nu}...\partial q^{\rho}} = f^{(n)}(q,p)(\epsilon_{\mu \xi}p^{\xi})(\epsilon_{\nu \eta}p^{\eta})... (\epsilon_{\rho \gamma}p^{\gamma})
\end{equation}
This constraint implies that $f(q,p)$ can be a function of the coordinates $q$ only
through $\eta \equiv \epsilon_{\mu \nu}q^{\mu}p^{\nu}$,
{\it i.e.} $f(p,q) = \tilde{f}(p,\eta)$.
With this starting point we can now consider the possibility
of $\lambda$-deforming (Snyder-deforming) such observables $\tilde{f}(p,\eta)$. On the basis of what we established above one concludes that the $\lambda$-deformed observable $\tilde{f}_\lambda(p,\eta)$ would have to admit description in terms of the
undeformed $\tilde{f}(p,\eta)$ as follows:
\begin{equation}
\tilde{f}_\lambda(p,\eta) \approx \tilde{f}(p,\eta) + \frac{\partial \tilde{f}(\eta, p)}{\partial \eta}\delta_\lambda \eta + \frac{\partial \tilde{f}(\eta, p)}{\partial p^{\nu}}
\delta_\lambda p^{\nu}
\nonumber
\end{equation}
where $\delta_\lambda \eta$ and $\delta_\lambda p^{\nu}$ are the $\lambda$-dependent
changes in $\eta$ and $p^{\nu}$ due to the deformation. For the
case here of interest, the Snyder deformation, we evidently have that $\delta_\lambda p^{\nu} =0$ (the representation (\ref{represnyder})
only involves a $\delta_\lambda q^{\nu} \neq 0 $). Moreover, evidently
 $\eta$ is just the classical Lorentz group generator (in a $1+1$-dimensional
 spacetime), and $\delta_{\lambda}\eta = 0$ since the Snyder spacetime has undeformed Lorentz symmetries.%on the basis of (\ref{represnyder}),
%{\it i.e.} $\delta_{\lambda} q^\mu \equiv x^\mu - q^\mu=XXXXXX$ we
%can also conclude that
%$$\delta_{\lambda} \eta = XXXXXXXXXXXXXXX= 0 \, .$$
Therefore in the Snyder case  all observables on the physical Hilbert space are
undeformed: $\tilde{f}_\lambda(p,\eta)=\tilde{f}(p,\eta)$.\footnote{There is an apparent loophole in this proof:
one might think that the invariant length scale, $\lambda$, available in  the 
theory here of interest could play a role in allowing to consider an apparently new class of observables, 
of the kind $\tilde{f}_\lambda(p, \eta) = \gamma(\lambda p) \tilde{f}(p,\eta)$. We feel that this is not a significant
limitation to our argument. Even without  $\lambda$ one could build such observables in terms of other available
  length/energy scales ({\it e.g.} the electron mass $m_e$ multiplied by a pure number $\alpha_\#$ such
  that $\alpha_\# m_e \equiv 1/\lambda$). Also notice that if we want to retain the transformation properties 
  of $\tilde{f}(p,\eta)$ we must have a scalar $\gamma(\lambda p)$, so that for a single-particle Hilbert space 
  (unless some preferred
   external momentum is introduced) $\gamma$ could be only a function of the mass $\gamma(\lambda m)$, {\it i.e.} a 
   quantity which takes a constant value over the whole physical Hilbert space.}

\section{Closing remarks}
We feel that the results here reported, besides evidently contributing
to the understanding of the much-studied Snyder noncommutativity,
provide a warning of broader potential technical relevance
and may also be relevant for a broader conceptual
assessment of the notion of ``empty spacetime".

We showed that Snyder's famous discretization of
spatial coordinates
is merely a property of ``empty spacetime", in the sense
that no trace of it can be found on the physical Hilbert space.
Is Snyder's deformation then irrelevant for physics? This is
what we established, but exclusively for free particles
in ``vanilla Snyder", {\it i.e.} the flat noncommutative spacetime
Snyder originally proposed. It will be interesting to check whether
the triviality of the deformation persists once interactions are
introduced. The most natural path for that would however take us out
of the framework here analyzed, since it should use quantum field theory. Also interesting is to contemplate the interplay between Snyder
noncommutativity
and spacetime curvature. While Snyder focused on the flat-spacetime
case, of course if his proposal has anything to do with Nature
it would have to be generalized to allow for gravity and curvature
of spacetime. This would in turn impose a reanalysis of one of the
key ingredient of the triviality we here exposed:
the integration measure (\ref{measure}), which was meaningful on
the kinematical Hilbert space but trivializes on the physical Hilbert
space,  only takes into account momentum-space curvature. If also spacetime
was curved a more complicated measure of integration should be introduced
to take that into account and the end result might not be trivial
even on the physical Hilbert space.

This notwithstanding, our findings  should invite prudence for other studies of quantum
geometries that remain confined on the kinematical
Hilbert space, without introducing on-shell particles.
Still, it should not be expected that all quantum properties of spacetime
introduced on the kinematical Hilbert space disappear on the physical Hilbert space of free particles.
What we found
in our previous study Ref.\cite{fuzzy2},
 concerning the so-called ``$\kappa$-Minkowski noncommutativity",
establishes that in some cases the implications of noncommutativity
of coordinates, introduced on the kinematical Hilbert space, does
affect significantly the structure of the physical
Hilbert space, even with just free particles. So this appears
to be an issue that requires a case-by-case analysis.

\bigskip
\bigskip

{\it This research was supported in part by
the John Templeton Foundation.}

%\newpage

\end{document}